\documentclass[11pt]{article}
\usepackage{amsmath}
\usepackage{amsfonts}
\usepackage{graphicx}
\usepackage{xcolor}
\usepackage{enumitem}
\definecolor{darkred}{rgb}{0.6,0,0}
\graphicspath{ {revision/} }

\begin{document}

\title{Chaitin's Omega and an Algorithmic Phase Transition}
\author{\\[10pt] Christof Schmidhuber\footnote{christof@schmidhuber.ch}\\[10pt]
	 \it{Zurich University of Applied Sciences, School of Engineering}\\	 [5pt]
	 \it{Technikumstrasse 9, 8401 Winterthur, CH-Switzerland}\\	 [10pt]}

\maketitle	

\begin{abstract}
We consider the statistical mechanical ensemble of bit string histories that are computed by a universal Turing machine. The role of the energy is played by the program size. We show that this ensemble has a first-order phase transition at a critical temperature, at which the partition function equals Chaitin's halting probability $\Omega$. This phase transition has curious properties: the free energy is continuous near the critical temperature, but almost jumps: it converges more slowly to its finite critical value than any computable function. At the critical temperature, the average size of the bit strings diverges. We define a non-universal Turing machine that approximates this behavior of the partition function in a computable way by a super-logarithmic singularity, and discuss its thermodynamic properties. We also discuss analogies and differences between Chaitin's Omega and the partition functions of a quantum mechanical particle, a spin model, random surfaces, and quantum Turing machines. For universal Turing machines, we conjecture that the ensemble of bit string histories at the critical temperature has a continuum formulation in terms of a string theory. 
\end{abstract}

\vspace{20pt}

{\small 
\hspace{-17pt}{\bf Keywords:}
Chaitin's Omega, Complexity, Turing Machine, Algorithmic Thermodynamics, Phase Transition, String Theory}

\newpage 
\setcounter{page}{2}

\section{Introduction}

In 1975, G. Chaitin \cite{C} introduced a constant associated with a given universal Turing machine $U$ \cite{T} that is often called the "halting probability" $\Omega$. It is computed as a weighted sum over all prefix-free input programs $p$ for $U$ that halt:	
\begin{equation}
\Omega_U=\sum_\text{halting p(U)} 2^{-l(p)}
=\sum_{l=1}^\infty N(l)\ 2^{-l}\label{A}
\end{equation}
where $p$ is a "program" (a bit string made up of $0$'s and $1$'s), $l(p)$ is its length (the number of bits), and $N(l)$ is the number of prefix-free programs of length $l$ for which $U$ halts. Turing machines and prefix-free bit strings are briefly reviewed in section 2 and in the appendix. For a general introduction to information theory, see \cite{V}.\\

Most of the discussion around $\Omega$ has focused on its first few digits, which are determined by the function $N(l)$ for small program length $l$. As every mathematical hypothesis can be translated into a halting problem (the question whether a given program halts for a given Turing machine), many long-standing mathematical problems could be solved if only one could compute $\Omega$ digit by digit. Unfortunately, $\Omega$ is not computable by any halting program, precisely because knowing $\Omega$ would imply that one could decide mathematical problems that are known to be undecidable in the sense of G\"{o}del's incompleteness theorem \cite{G}. Moreover, even when the first few digits of $\Omega$ are computable for a given universal Turing machine $U$, they are not universal: they depend on the choice of $U$.\\
	
In this note, we will therefore not be concerned with the contribution of short programs to $\Omega$, nor will we dwell much on the issues of incompleteness and undecidability. Instead, we will focus on the contribution of very long programs to $\Omega$, i.e., on the behaviour of $N(l)\cdot2^{-l}$ as $l\rightarrow\infty$. More precisely, following \cite{Tad02,calude,baez}, in a generalization of (\ref{A}), we consider the statistical mechanical ensemble of bit string histories with partition function	
\begin{equation}
Z_U(\beta)=\sum_\text{halting p(U)} \exp\{-\beta\cdot l(p)\}
\text{\ \ \ \ \ \ with\ \ \ \ \ \ } 
\beta={1\over{kT}},\label{B}
\end{equation}
where $k$ is the Boltzmann constant and $T$ the temperature as usual in statistical mechanics (see, e.g., \cite{Z} for a review of statistical mechanics and field theory). We will study $Z$ as a function of $\beta=\beta_c+\epsilon$ in the vicinity of the "Chaitin point" $\beta_c=\ln 2$ with $\epsilon\ll 1$ (assuming a binary alphabet; for general Turing machines, $\beta_c$ is the log of the alphabet size).\\

We find that the "Chaitin point" $\beta=\beta_c=\ln 2$ corresponds to a critical temperature, at which a first-order phase transition occurs. This phase transition has very curious properties. In particular, the free energy is almost discontinuous: it converges more slowly to its finite critical value than any computable function. However, in a finite universe with limited computation time, this effective discontinuity is invisible. We illustrate this type of transition in a toy model, namely a non-universal Turing machine (the "counting machine") that approximates this behavior of the free energy by a super-logarithmic singularity.\\

At the critical point, the average size of the output bit strings diverges. This leads us to the fascinating question whether there is a continuum formulation of our bit string ensemble at the Chaitin point in terms of some string theory \cite{S}, in which the two-dimensional string world-sheet is spanned by the bit string and the computation time. To set the stage for such a link between algorithmic information theory and two-dimensional field theory, we point out formal analogies between ensemble (\ref{B}) at the critical point and (i) a quantum mechanical relativistic particle, (ii) the condensation of defects in a spin model, and (iii) the theory of random surfaces. We also interpret the ensemble (\ref{B}) as a probabilistic Turing machine and show how it can be extended it to a quantum Turing machine.  

\section{Relation to Previous Work}

The generalization (\ref{B}) of Chaitin's halting probability has previously been studied by Tadaki \cite{Tad02}, who investigated the degree of randomness of the real number $Z_U(\beta)$, written in binary form. The relation with statistical mechanics was pointed out by Calude and Stay \cite{calude}, who also discuss variants of (2), in which the sum runs over general (as opposed to only prefix-free) programs (the partition function then diverges at $\beta=\ln2$, instead of converging to Chaitin's $\Omega$). The statistical mechanical approach was formulated more mathematically by Tadaki in \cite{Tad08}. \\

Baez and Stay \cite{baez} define the corresponding "algorithmic" versions of the specific heat and other thermodynamic quantities. Morevover, they formally extend the Gibbs factor (\ref{B}) by including two other terms, corresponding to the logarithm $E(p)$ of the computation time of the Turing machine, and to the expectation value $N(p)$ of the output bitstring string (interpreted as a natural number in binary form):
\begin{equation}
\exp\{-\beta_1\cdot l(p)-\beta_2\cdot E(p)-\beta_3\cdot N(p)\}.\label{B1}
\end{equation}

Algorithmic versions of the Carnot cycle are also discussed in \cite{baez}, where it is pointed out that the partition function has a singularity at $\beta_1 = \ln 2,\ \beta_2=\beta_3=0$. Tadaki \cite{Tad12,Tad12b} discuss computational aspects of this "algorithmic phase transition".\\

Our paper complements this previous work by studying the nature of this algorithmic phase transition from a more physical point of view. In particular, a key question about phase transitions is, whether they are first-order or second-order. As mentioned, we resolve this in sections 7 and 8 by showing that this one is an exotic first-order transition. \\

The sum (\ref{B}) over $p$ can be re-written in terms of a sum over all output bit strings $B$:
\begin{eqnarray}
Z_U(\beta)&=&\sum_\text{B} \exp\{-\beta\cdot K_\beta(B)\}\label{B21}\\ 
\text{with}\ \ K_\beta(B)&=&-{1\over\beta}\ln  \sum_{p(B)} \exp\{-\beta\cdot l(p)\}\ \rightarrow\ \ K(B)\ \ \text{as}\ \ \beta\rightarrow\infty.\label{B2}
\end{eqnarray}
Here, the second sum runs over all programs $p(B)$ that halt and produce $B$. $K(B)$ is the Kolmogoroff complexity, i.e., the length of the shortest program that makes $U$ compute $B$ \cite{kol}. For general bit strings, $K(B)$ is not computable by any halting program. At the critical point $\beta=\beta_c$, $\exp \{ -\beta_c K_{\beta_c}(B)\}$ is the algorithmic Solomonoff probability \cite{Sol}. \\

According to Levin's coding theorem \cite{Lev}, $K_{\beta_c} (B)\le K(B)+c$ with a constant $c$ that does not depend on $B$. We also have $K(B)<K_{\beta_c}(B)$, because the sum in (\ref{B2}) runs over all programs that produce the output bit string $B$, not only the shortest one. 
In recent work, Kolchinsky and Wolpert \cite{Wol} interpret the differences
$$C(B)=K_{\beta_c}(B)-K(B)\ \ ,\ \ Q(p(B))=l(p(B))-K(B)$$
in terms of the minimum heat that is released into the environment by the computation in the sense of \cite{lan}. In \cite{Wol}, properties of this generated heat are discussed both for the "coin flipping realization" (\ref{A}) of the Turing machine $U$ and for other realizations.\\

In a separate line of work (see \cite{Y1,Y2,Y}, and references therein), Manin considers a self-adjoint Hamiltonian $H$ acting on the Hilbert space spanned by the output bit strings $\vert B\rangle$, viewed as states. If $\beta=it$ is interpreted as imaginary time, the operator $\exp\{iHt\}$ describes a unitary time evolution. $H$ is defined such that $\vert B\rangle$ are its eigenvectors with eigenvalues $K(B)$. Manin relates this model to error-correcting codes, to Zipf's law, and to renormalization in field theory (I thank D. Murfet for pointing this out to me). \\

Inspired in part by this work, in section 9 we formulate ensemble (\ref{B}) as a probabilistic Turing machine that runs in a different time $\tau$, defined as the position of the head on the read-only program tape. Starting with a blank work tape, the final state is a superposition of output bit strings. 
While this time evolution is irreversible, and therefore the transfer matrix is not unitary, it can be extended to act on halting states such that it asymptotically approaches Manin's unitary time evolution, as well as generalizations thereof, after an initial phase of heat production.\\

One of the many other interesting issues that have been addressed in the context of Turing machines and Hamiltonian formulations is that in some cases it is an undecidable problem whether or not the energy spectrum has a vanishing mass gap \cite{cub,cub2}.

\section{Turing Machines and Prefix-free Programs}

We follow Chaitin's definition of a Turing machine, which is reviewed in appendix A. A bit string $y$ is called a prefix of a bit string $x$, if $x$ can be written as a concatenation $x=yz$, with a third bit string $z$. A set of bit strings is called prefix-free, if no bit string is a prefix of another. For the current argument, it is sufficient to think of a Turing machine $T$ as a map ("computation") from a set $P$ of prefix-free input bit strings $p\in P$ ("programs"), for which the computation halts, to the set $O$ of arbitrary output bit strings of any length:
$$T: p\in P\rightarrow T(p)\in O$$

The output bit strings are written on a "work tape" that extends infinitely in both directions. The computation manipulates them until it halts. The prefix-free input bit strings are written on a finite read-only "program tape" (see appendix A for details). The prefix-free input programs $p$ are what we sum over in (\ref{B}), and whose lengths $l$ play the role of the energy in the Boltzmann factor.

\begin{figure}[h]\centering
	\includegraphics[height=3.2cm]{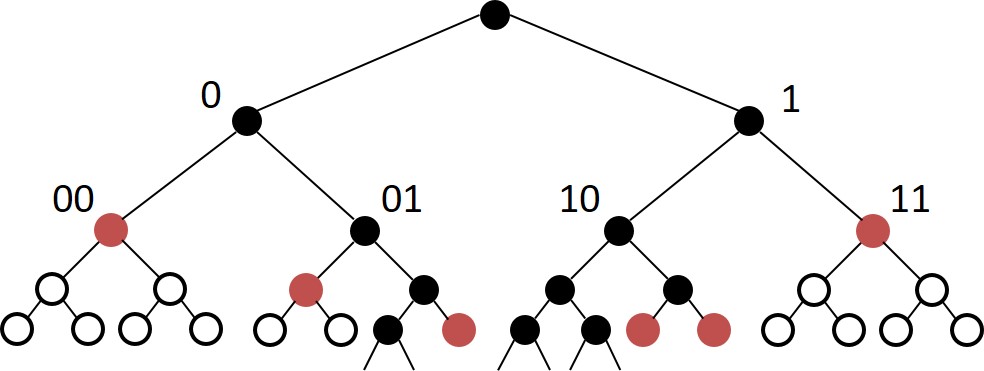} 
	\caption{Tree representation of prefix-free bit strings}
	\label{figA}
\end{figure}

One may represent a set $X$ of prefix-free bit strings by a tree (fig. \ref{figA}). The vertices in the $(l+1)$-th line (or $l$-th generation) of the graph represent the $2^l$ binary numbers $b_l$ with $l$ digits. Branching to the left appends a $0$, branching to the right appends a $1$ at the end of $b_l$ to yield the next generation of $b_{l+1}$. At each vertex, the corresponding number $b_l$ is either added to the set $X_l\subset X$ of prefix-free programs $p_l$ of size $l$ (red dots) or not (black dots). Black dots are prefixes (parents, grand-parents, ...) of red dots and give birth to two children; we assume that each black dot is the prefix of at least one red dot. Red dots have no children. In the figure, white dots represent bit strings that are never born. \\

Let $n_l$ be the number of red dots (prefix-free programs) of length $l$. Let $m_l$ be the number of black dots (prefixes) of length $l$. Let $w_l=2^l-n_l-m_l$ be the number of white dots of length $l$. We define the percentages $Q_l=w_l\cdot 2^{-l}$ of white dots and $P_l=n_l\cdot 2^{-l}$ of red dots in the $l$-th generation and get 
\begin{equation}
P_l=Q_{l+1}-Q_l\ \ \ \text{with}\ \ \ \lim_{l\rightarrow\infty}Q_l=\sum_{l=1}^\infty P_l =\sum_{p_l\in X} 2^{-l} =1,\label{H}
\end{equation}
where the last equation states that "Kraft's inequality is satisfied with equality" (see \cite{V}).
As an example of a set of prefix-free programs, consider "Fibonacci coding": a child is a member of $X$, if its last $2$ digits are "1" or - in a slight generalization - if its last $N$ digits are "1". In this case, one easily verifies that $P_l$ falls off exponentially as $l\rightarrow\infty$.\\

For a given Turing machine $T$, there are two kinds of red dots: ${\tilde n}_l$ halting programs and $n_l-{\tilde n}_l$ non-halting programs. We denote by $h_l={\tilde n}_l /n_l$ the fraction of programs in the $l$-th generation that halt. Then the partition function (\ref{B}) can be written as
\begin{equation}
Z_U(\beta)=\sum_{l=1}^\infty P_l\cdot h_l\ e^{-\epsilon\cdot l}\ \ \ \ \text{with}\ \ \ \epsilon=\beta-\beta_c\ ,\ \beta_c=\ln 2.\label{BC}
\end{equation}

\section{Universality of the Singularity}

At the critical point $\beta=\beta_c=\ln 2$, our partition function (\ref{BC}) is Chaitin's $\Omega$ (\ref{A}):
$$
Z_U(\beta_c)=\sum_{l=1}^\infty P_l\cdot h_l =\Omega<1 \label{BB}
$$

For $\beta<\beta_c$, the partition function diverges, as long as $P_l\cdot h_l$ falls off more slowly than exponentially as $l\rightarrow\infty$, which is the case for any universal Turing machine $U$ (see below). How exactly does $Z_U$ approach $\Omega$ as $\beta$ approaches $\beta_c$ from above? Let us first discuss in how far this singularity near $\beta=\beta_c$ is universal, i.e., independent of $U$.\\

A universal Turing Machine ("UTM") $U$ is one that can simulate any other Turing machine $T_i$ in the following sense: there is a finite bit string ("translator program") $c_i$ such that for each program $p$, $U(c_i p)=T_i(p)$. I.e., if $p$ makes $T_i$ compute an output bit string, the concatenation $c_i p$ makes $U$ compute the same output bit string. Let $C_i$ be the finite length of the program $c_i$. Then the partition function $Z_U(\beta)$ of the UTM contains the partition function $Z_i (\beta)$ of $T_i$ as a subset:
$$Z_U(\beta)\ge e^{-\beta C_i}\cdot Z_i (\beta)$$

This applies to all Turing machines $T_i$. Thus, as $\epsilon=\beta-\beta_c\rightarrow0$, the Turing machine $T_i$ with the strongest singularity (i.e., with the largest derivative $Z_i'(\epsilon)$ at $\epsilon\sim0$) dominates the singularity of the partition function $Z_U(\beta)$ at $\beta=\ln 2$. As this applies to all $U$, we conclude that this singularity is universal, i.e., independent of the choice of the UTM, up to an overall pre-factor $2^{-C_i}$.\\

Our ensemble (\ref{B}) includes only programs that halt. This makes it intractable, as it is generally an undecidable question whether a given Turing machine halts for a given program. Thus, the factor $h_l$ in (\ref{BC}), Chaitin's $\Omega$, and the partition function $Z_U(\beta)$ are actually not computable by any halting program. These issues around un-decidability and non-computability, fascinating as they may be, will not play a major role here. It is clear from (\ref{BC}) that the strongest singularity in $\epsilon$ corresponds to the product $P_l\cdot h_l$ that decays most slowly as $l\rightarrow\infty$. Thus, the non-computable factor $h_l$ can only make this singularity weaker. We will therefore first discuss non-universal Turing machines $T_i$ for which all programs halt (i.e. $h_l=1$), and then return to universal Turing machines towards the end.\\

As an example of a function $P_l$ that converges more slowly than that from Fibonacci coding, let the $N$ of Fibonacci coding grow with the program length: $N(l)=\text{int}(1+\lg l)$, where $\lg\equiv \log_2$. In this case, it is not difficult to see that $P_l$ decays like a power of $l$:
$$P_l\propto l^{-\alpha}\ \ \text{as}\ l\rightarrow\infty\ \ \text{with}\ \alpha>1\ \ \ \Rightarrow\ \ \ 1-Z(\epsilon)\propto \epsilon^{\alpha-1} \ \ \text{as}\ \ \epsilon=\beta-\beta_c\rightarrow 0    $$ 
Next, we present a machine that yields a much stronger, super-logarithmic singularity.

\section{The Counting Machine}

We now define a Turing machine $T_0$ that we call the "counting machine", corresponding to a particular set of prefix-free programs, that always halts. We will then show that its partition function (\ref{BC}) has a computable, super-logarithmic singularity that, for our purposes, serves as a good model of the singularity of UTM's.\\

Let us first describe the output of the machine $T_0$. Given any infinite input bit string $p$ on the program tape, $T_0$ writes a number $N$ of $1$'s in a row on its otherwise blank work tape and then halts. We call $p_N$ the prefix of $p$ consisting only of those bits of $p$ that have been read by the time the machine halts, i.e., the machine halts on the last bit of $p_N$. This defines a set $P$ of prefix-free input programs $p_N\in P$. We will construct $T_0$ such that any number $N\in\mathbb{N}_0$ of 1's appears as the output bit string of exactly one such $p_N$, namely:
\begin{eqnarray}
\text{for } N<3: && p_0=00\ ,\ p_1=01\ ,\ p_2=10\ \ \ \text{with length}\ \ \ l_N=2\notag\\
\text{for } N=3: && p_3=110\ \ \ \text{with length}\ \ \ l_N=3\label{J}\\
\text{for } N>3: && p_N=11 n_2 ... n_k N 0\ \ \ \text{with length}\ \ \ 
l_N=6+n_2+...+n_{k}\notag
\end{eqnarray}
where $n_k$ is the binary length of $N$, $n_{k-1}$ is the binary length of $n_k$, and so on, until a length $n_1=3=11_2$ is reached. For $N>2$, $p_N$ begins with "11" and ends with "0". The number of iterations $k$ can be recursively expressed as follows:

\begin{equation}
k(N)= \left\{ \begin{array}{l}
0\ \text{if}\ N<4\\
1+k(1+\lg N)\ \text{if}\ N\ge4 \end{array}\right. 
\label{JJJ}\end{equation}\

This yields, e.g., $k(4)=k(7)=1, k(8)=k(127)=2, k(128)=3$, and so on. \\

Next, we describe how $T_0$ reconstructs $N$ from $p_N$. Given an infinte string $p$ on the program tape, $T_0$ proceeds as follows:
\begin{enumerate}
\item $T_0$ reads the first two digits $n_1=p_1 p_2$ of $p$. If $n_1=00_2$, $T_0$ leaves the work tape blank and halts; if $n_1=01_2$, $T_0$ writes $1$ on the work tape and halts; if $n_1=10_2$, $T_0$ writes $11$ on the work tape and halts. If $n_1=11_2$, $T_0$ reads the next digit $p_3$ and defines the new integer $m_1=3$. 
\item If $p_{m_1}=p_3=0$, $T_0$ writes $1^{n_1}=111$ on the work tape, then halts. If $p_{m_1}=p_3=1$, $T_0$ reads the next $n_1=3$ digits $p_{m_1+1},..,p_{m_1+n_1}$ of $p$, i.e., $p_4,p_5,p_6$. $T_0$ defines $m_2=m_1+n_1=6$ and $n_2=p_3 p_4 p_5$ (the concatenation with $p_3$ but without $p_6$)
\item[\vdots]
\item[i.] In the $i$-th step, if $p_{m_{i-1}}=0$, $T_0$ writes $1^{n_{i-1}}$ on the work tape and halts. If $p_{m_{i-1}}=1$, $T_0$ reads in the next $n_{i-1}$ digits. It defines $m_i=m_{i-1}+n_{i-1}$ and the concatenation $n_i=p_{m_{i-1}}\ ...\ p_{m_i-1}$ and moves on to step $(i+1)$, until $T_0$ halts. If $T_0$ halts in the $i$-th step, then $i$ is related to $k$ of (\ref{JJJ}) by $k=i-2$, and $N=n_{i-1}$.
\item[] E.g., in the third step, $p_3=1$ and $m_2=6$. Suppose, $n_2=p_3p_4p_5=101_2=5$. If the sixth digit $p_6$ of $p$ is 0, $T_0$ writes a sequence of $n_2=5$ 1's on the tape and halts. In this case, $k=1$ and $N=5$. However, if $p_6=1$, $T_0$ reads in the next five digits $p_7...p_{11}$, defines $m_3=m_2+n_2=11_{10}$ and $n_3=p_6...p_{10}$, and moves on to step 4.
\end{enumerate}

As an example, consider the input bit string $11100110100$. Then $n_1=11_2$, so in step $2$, $T_0$ defines $m_2=6,n_2=100_2=4$. In step $3$, since $p_6=1$, $T_0$ sets $m_3=10$ and reads in the 4-digit number $n_3=1101_2=13$. In step $4$, since the next digit $p_{10}$ is a $0$, $T_0$ writes $N=13$ digits $1$ in a row and  halts. Only the first 10 digits $1110011010$ of the input bit string constitute an element of $P$. More generally, if the counting machine halts in step $k$, the first $m_{k-1}$ digits of the input string constitute an element of $P$. The first elements are:
$$P=\{00, 01, 10, 110, 111000, 111010, 111100, 111110, 1110010000, ... \}$$

One may verify that any number $N$ of $1$'s in a row appears as the output bit string of exactly one program $p_N\in P$, as claimed above. It is also clear that $P$ is complete in the sense that it cannot be enlarged by any additional bit string without spoiling its property of being prefix-free. As a result, (\ref{H}) implies that $Z(\beta_c)=1$.\\

Although the counting machine $T_0$ only produces bit strings that are trivial in the sense that they contain only 1's, variants of the counting machine can be used to make them less trivial in subsequent steps. E.g., in appendix A6, a variant is discussed that generates all integers $m$ in binary form, such as $m=20=10100_2$, and then overwrite the 1's by repeating $m$ until the bit string ends: "1010010100...". Another variant generates all integers $K$, and then create $K$ "kinks" on the bit strings resulting from step 2, by flipping all bits after certain points on the strings. Thus, the counting machine is also a useful tool for systematically generating nontrivial output bit strings of increasing complexity.\\

More generally, the counting machine $T_0$ can be used whenever one needs a highly compact specification of large numbers $N$ by prefix-free programs. Of course, other sets of prefix-free programs may give a shorter description of {\it individual} large numbers, such as $2^{2^{1024}}$, at the expense of the {\it average} large number.\\

Appendix A4 presents a concrete implementation of the counting machine $T_0$.

\section{Super-logarithmic Singularity}

In this section, we compute how the partition function (\ref{BC}) approaches its critical value as $\beta$ approaches $\beta_c$ from above in the case of the counting machine. The counting machine halts for every input program ($h_l=1$) and therefore has a computable partition function 

\begin{equation}
\hat Z(\beta)=\sum_\text{all p} \exp\{-\beta\cdot l(p)\}
= \sum_{k=0}^\infty \hat Z_k(\beta) 
\text{\ \ \ \ \ \ with\ \ \ \ \ \ } \hat Z(\beta_c)=1,\label{sum}
\end{equation} 
where $\hat Z_k(\beta)$ is the contribution from programs $p$ that halt after $k$ iterations, $k$ being defined in (\ref{JJJ}). Using (\ref{J}), we expand:
\begin{eqnarray}
\hat Z_0(\beta)&=&3 e^{-2\beta} + e^{-3\beta}\ \ ,\ \ \hat Z_1(\beta)\ =\ 4 e^{-6\beta}\notag\\ 
\hat Z_2(\beta)&=&8 e^{-10\beta} + 16 e^{-11\beta} +32 e^{-12\beta}+64 e^{-13\beta}\notag\\
\hat Z_k(\beta)&=&\sum_{n_2,...,n_k,N} e^{-\beta\cdot(6+n_2+...+n_{k})}\notag\\
&\sim&\ \ \sum_{n_2,...,n_{k}} 2^{-(6+n_2+...+n_{k-1})}\cdot {1\over2}\ e^{-\epsilon n_k}\ \ \ \text{with}\ \ \ \epsilon=\beta-\beta_c,
\label{CO}\end{eqnarray}
where $n_2$ runs from 4 to 7, $n_{i+1}$ runs from $2^{n_i-1}$ to $2^{n_i}-1$, and $N$ runs from $2^{n_k-1}$ to $2^{n_k}-1$. In the last line, we have expanded near $\beta=\beta_c=\ln 2$, and kept only the leading term in $\epsilon$, noting that $n_k\gg n_{k-1}$. For a given $k$, let $\Lambda_k$ be the largest possible value of $n_k$:
\begin{equation}
\Lambda_1=3 ,\ \ \ \Lambda_2=7,\ \ \    \Lambda_3=127,\ \ \    \Lambda_{k+1}=2^{\Lambda_k}-1.  
\label{F}\end{equation}

If $\Lambda_{k-1}\gg1/\epsilon$, we can approximate $\hat Z_k$ in (\ref{CO}) by 0, since the minimum value of $n_k$ is $\Lambda_{k-1}+1$. On the other hand, if $\Lambda_{k}\ll1/\epsilon$, we can approximate $\epsilon$ by $0$ in $\hat Z_k$. This yields $\hat Z_0=7/8,\hat Z_1=1/16$. Noting that there are always  $2^{n_i}/2$ possible values for $n_{i+1}$, in the case $\Lambda_{k}\ll1/\epsilon$ we can iteratively perform the sum over $n_2,...,n_{k}$ for $k>1$ to obtain
$$
\hat Z_{k}={1\over 2}\sum_{n_2,...,n_{k}} 2^{-6-n_2-...-n_{k-1}}={1\over 4}\sum_{n_2,...,n_{k-1}} 2^{-6-n_2-...-n_{k-2}}  =...={1\over 2^{k+3}}  
$$
We now perform the sum (\ref{sum}) over $k$ and first consider the (rare) case where $1/\epsilon =\Lambda_K$ for some $K$. In appendix A5, it is shown that, in this case, $\hat Z_K=2^{-K-3}, Z_{K+1}=0$ to high accuracy already for $K\ge4$. Thus, 
\begin{eqnarray}
\hat Z(\epsilon)={7\over8}+ {1\over16}+ \sum_{k=2}^{K}\hat Z_k= 1-{2^{-K-3}}
\ \ \ \text{with}\ \ \ {1\over\epsilon}=\Lambda_K
\label{LLL}\end{eqnarray}
The singularity in $\epsilon$ comes from the dependence of $K$ on $\epsilon$. To continue (\ref{LLL}) to general $\epsilon$, we use the "super-logarithm" $\text{slog}_2(x)$ with basis $2$ in the so-called "linear approximation":
\begin{eqnarray}
&&\text{slog}_2(x)=\left\{ \begin{array}{l}
x-1\ \ \ \text{if}\ 0< x\le1\\
\text{slog}_2(\lg(x))+1\ \ \ \text{if}\ x>1 \end{array}\ \right.
\label{MMM}\end{eqnarray}
Its integer values are $\text{slog}_2(1)=0, \text{slog}_2(2)=1, \text{slog}_2(4)=2, \text{slog}_2(2^x)= \text{slog}_2(x)+1$. Real values of $\text{slog}_2$ are interpolated from its integer part $\lg^-(x)=\text{int}(\text{slog}_2(x))$ by
\begin{eqnarray}
\text{slog}_2(x)=\lg^-(x)+\lg ... \lg x\ \ \ \text{with}\ 1+\lg^-(x)\ \text{iterations}
\notag\end{eqnarray}
We can now express $K(\epsilon)$ in terms of the super-logarithm by noting from (\ref{F}) that $\text{slog}_2(\Lambda_{k+1})\rightarrow \text{slog}_2(\Lambda_k)+1$ to very high accuracy already for $k>2$:
$\text{slog}_2(\Lambda_1)=1+\lg\lg 3\sim 1.66 ,\ \   \text{slog}_2(\Lambda_2)=2+\lg\lg\lg 7\sim 2.57, \ \   \text{slog}_2(\Lambda_k)=k+0.57$
\begin{eqnarray}
\Rightarrow\ \ K(\epsilon)&\sim& \text{slog}_2({1/\epsilon})-\phi\ \ \ \ \text{with}\ \ \ \phi=0.57...\notag\\
\hat Z(\epsilon)&\sim&1-\lambda\cdot 2^{-\text{slog}_2(1/\epsilon)} 
= 1-\lambda\cdot 2^{-\lg^-(1/\epsilon)}\cdot \{\lg ... \lg {(1/\epsilon)}\}^{-1} ,   \label{G}
\end{eqnarray}
where $\lambda=2^{\phi-3}\sim 0.186$, and there are $\lg^-(1/\epsilon)$ iterations of the logarithm in the last line. \\

This continues (\ref{LLL}) to any $\epsilon$. Although the continuation (\ref{MMM}) of the super-logarithm to real values, and thus the continuation (\ref{G}) of $\hat Z(\epsilon)$, is not unique, different continuations differ only by sub-leading orders in $\epsilon$. Thus, (\ref{G}) is the leading singularity of the partition function $\hat Z(\epsilon)$ at the critical point. This partition function is plotted in  fig. \ref{figC}. It converges extremely slowly to $1$ as $\epsilon\rightarrow0$, and is continuous but "almost" discontinuous. 

\begin{figure}[h]\centering
	\includegraphics[height=5cm]{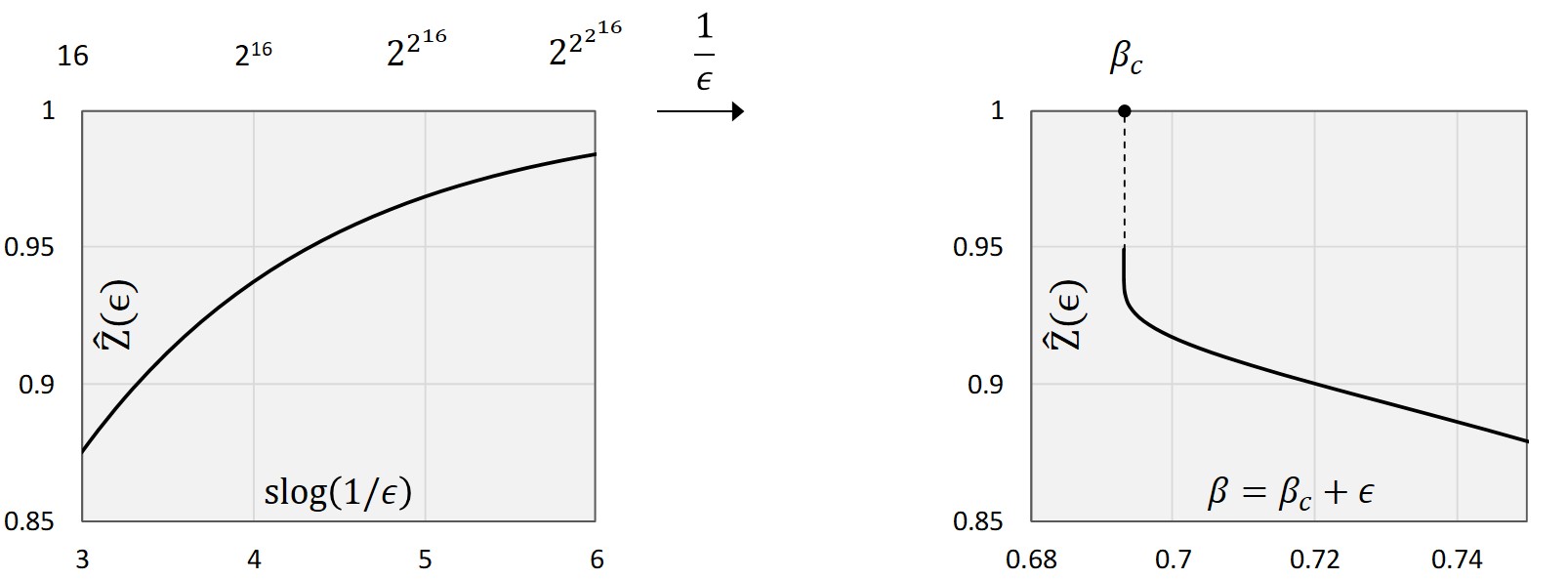} 
	\caption{$\hat Z(\epsilon)$ as a function of $1/\epsilon$ (left) and $\beta$ (right)}
	\label{figC}
\end{figure}

\section{Critical Behavior}

Armed with the results of section 5, let us now examine the phase transition for the counting machine near the critical point $\beta=\beta_c+\epsilon$ with $\beta_c=\ln 2, \epsilon\ll1$. The free energy $F$ is:
\begin{eqnarray}
\hat Z(\beta)=e^{-\beta F}=\sum_p e^{-\beta l(p)} \ \  \Rightarrow\ \ F(\beta)=-{1\over\beta} \ln \hat Z(\beta)\label{countZ}
\end{eqnarray}
In our ensemble, the program length plays the role of the energy with expectation value
\begin{eqnarray}
\langle l\rangle_\beta =-\partial_\beta  \ln \hat Z(\beta)\label{countY}
\end{eqnarray}
The heat capacity is (using $T \partial_T=-\beta\partial_\beta$)
$$C(T)=-T {\partial^2 F\over\partial T^2 } \ \ \sim\ \ -\partial_\beta \ln \langle l\rangle\ \ +\ \ \text{higher orders in }\epsilon$$

According to the Ehrenfest classification, in a zeroth-order phase transition the free energy $F(T)$ is discontinuous at a critical point $T=T_c$. In a first-order transition, $F(T)$ is continuous but $\partial_T F(T)$ is discontinuous, the gap being the latent heat. In a second-order transition, $\partial_T F(T)$ is also continuous, but some higher-order derivative of $F(T)$ is discontinuous \cite{Z}.
In our case, 
$$\hat Z(\epsilon)=1- \lambda\cdot 2^{-\text{slog}_2(1/\epsilon)} \ =\ 1-\lambda\cdot 2^{-\lg^-(1/\epsilon)}\cdot \{\lg ... \lg {(1/\epsilon)}\}^{-1}
$$
where $\lambda\sim 0.186$,  $\lg^-$ is the integer part of the super-logarithm and we have $\lg^-(1/\epsilon)$ iterations of the logarithm. Thus, in the limit $\epsilon\rightarrow0$, we have, to leading order in $\epsilon$:
\begin{eqnarray}
F(\epsilon)\ \ &\propto& \ \ -\lambda\cdot 2^{-\lg^-(1/\epsilon)}\cdot \{\lg ... \lg {(1/\epsilon)}\}^{-1}\notag\\  
\langle l\rangle(\epsilon)\ \ &\propto&\ \ [\ \epsilon \cdot \lg {1\over\epsilon}\cdot\lg \lg {1\over\epsilon}\cdot...\cdot(\lg ... \lg {1\over\epsilon})^2\ ]^{-1} \label{thermo} \\
C(\epsilon)\ \ &\propto&\ \ [\ \epsilon^2 \cdot \lg {1\over\epsilon}\cdot\lg \lg {1\over\epsilon}\cdot...\cdot(\lg ... \lg {1\over\epsilon})^2\ ]^{-1}\notag  
\end{eqnarray}

$F(\epsilon)$ is finite at the critical point. It is continuous, but almost discontinuous. Thus, the phase transition is first-order, but almost zeroth order. We also see that the latent heat is infinite, meaning that the average program size $\langle l\rangle$ diverges at the critical point. The average size $N$ of the output strings also diverges, as $l\sim\lg N+\lg\lg N +...$ \\

To put things into perspective, the age of the universe, as measured in Planck times, is $T_u\sim2^{200}$. For $\epsilon< 1/T_u$, one needs to consider input bit strings of length $l>T_u$, and therefore computation times $>T_u$, to compute $F$ and $\hat Z$ at $\beta=\beta_c+\epsilon$. The super-logarithm of $T_u$ is about 4.6, so for $\epsilon$ of order $2^{-200}$, $\hat Z$ is still about $0.76\%$ away from 1. To get a super-logarithm of 5, we would need a universe of age $2^{65'536}$ Planck times. Even then, $\hat Z$ would still be $0.58\%$ away from 1. In this sense, the super-logarithmic singularity (the dashed line in fig. 2) is invisible at least for all bit string ensembles that can be computed in our universe, and  $\text{slog}_2(1/\epsilon)$ is effectively cut off at $4.5-5$. 

\section{Singularity for Universal Turing Machines}

In the previous section, we have discussed the singularity of the partition function (\ref{BC}) near $\epsilon=0$ for the non-universal counting machine. How does it compare with the singularity for a universal Turing machine?\\

Since it is generally an undecidable question whether a given Turing machine halts for a given program, for a UTM the function $h_l$ in (\ref{BC}) and Chaitin's $\Omega$ are not computable by any halting program. Neither is the singularity of $Z(\epsilon)$ at the critical point computable. In fact, $Z(\epsilon)$ converges towards $\Omega$ more slowly than any computable function.

To see this, let us slightly modify the last step of the counting machine of section 4: if, in the $i$-th step, $p_{m_{i-1}}=0$, the modified $T_0$ switches into a new mode: instead of writing $1^{n_{i-1}}$ on the work tape, it reads the next $\Sigma(n_{i-1})$ digits of the program $p$ from the program tape, where $\Sigma(n)$ is the busy-beaver function. The modified machine $\tilde T_0$ writes those digits on the work tape and then halts. Formula (\ref{CO}) thus gets replaced by 
$$
\tilde Z_k(\beta)=\sum_{n_2,...,n_k}\sum_{N=0}^{2^{\Sigma(n_k)}} e^{-\beta\cdot(6+n_2+...+n_{k}+\Sigma(n_k))}
\sim\sum_{n_2,...,n_{k}} 2^{-(6+n_2+...+n_{k})}\cdot {1\over2}\ e^{-\epsilon\cdot \Sigma(n_k)}\ \ \ \label{CD}
$$\
$\Sigma(n)$ is known to diverge faster than any computable function as $n\rightarrow\infty$. This implies that $\tilde Z(\epsilon)$ converges more slowly than any computable function to its critical value $1$ for the modified machine $\tilde T_0$. Now, any UTM $U$ simulates the modified machine $\tilde T_0$, if it is fed with all possible input programs. This implies that, for any UTM, $Z_U(\epsilon)$ converges more slowly than any computable function to its critical value $\Omega$. \\

The conclusions for UTM's are thus similar as for the counting machine: first of all, at the critical point, the phase transition is first-order but almost zeroth-order, with a divergent latent heat (i.e. average program size). The average size of the output bit strings also diverges, as $U$ simulates $T_0$, among other machines. This divergent average program size is a pre-condition for a potential continuum limit of (\ref{B}) at $\beta=\beta_c$, where very long bit strings might be described by continuous strings. This will be further discussed below. \\

Second, strictly speaking, $Z_U(\epsilon)$ for a UTM is continuous at $\epsilon=0$, but in practise, its behavior is indistinguishable from a discontinuity. This seems to make it impossible to get a better understanding of the Chaitin point (\ref{A}) by first studying the ensemble (\ref{B}) at low temperature (high $\beta$), and then taking the limit $\beta\rightarrow\beta_c$. \\

However, as in the case of the counting machine, this discontinuity is invisible in finite computation time. Thus, we can at least interpolate between the low-temperature phase and a "real-world" version of the Chaitin point (\ref{A}), in which the computation time is limited by the age $T_u$ of the universe. Alternatively, we could limit the length $L$ of the output bit strings $B$ to $L\le T_u$ (roughly the diameter of the observable universe), and count only the shortest program that produces each $B$. As $B$ is produced by at least one program of size $L+O(1)$, namely "print $B$", this also limits the computation time. For most $B$, this amounts to replacing $K_\beta(B)$ by the Kolmogoroff complexity $K(B)$ in (\ref{B21},\ref{B2}). 

\newpage
\section{Analogies with Quantum Mechanics}

In this section, we discuss analogies and differences of the statistical-mechanical model (\ref{B}) with quantum mechanics, both in the path integral- and the Hilbert space formulation. Our goal is to use these methods to better understand the approach to the critical point.\\

So far, we have only considered the input programs $p$ in (\ref{B}), and not the bit strings $b$ that they produce on the work tape during the computation. Let us now rewrite (\ref{B}) as a sum over paths in the space of configurations of the Turing machine. At any point in time, a configuration is uniquely specified by the 3-tuple \cite{fort}
$$S=\{b, s, k\},$$
where $b$ is the bit string on the work tape, $s$ is the state of the Turing machine, and $k$ is the position of the head. $k=1$ means that the head sits on the first non-blank bit of the work tape. We define "program time" $\tau\in\{0,1,2,...,l\}$ as the position of the head on the read-only program tape. Let us represent a computation by a path $S(\tau)$ over the configurations at the time the $\tau$-th input bit is first read in, i.e, immediately before the path splits. The path ends at $\tau=l, b(l)=B, s(l)=$"Halt", where $B$ is the output string. \\

\begin{figure}[h]\centering
	\includegraphics[height=5.5cm]{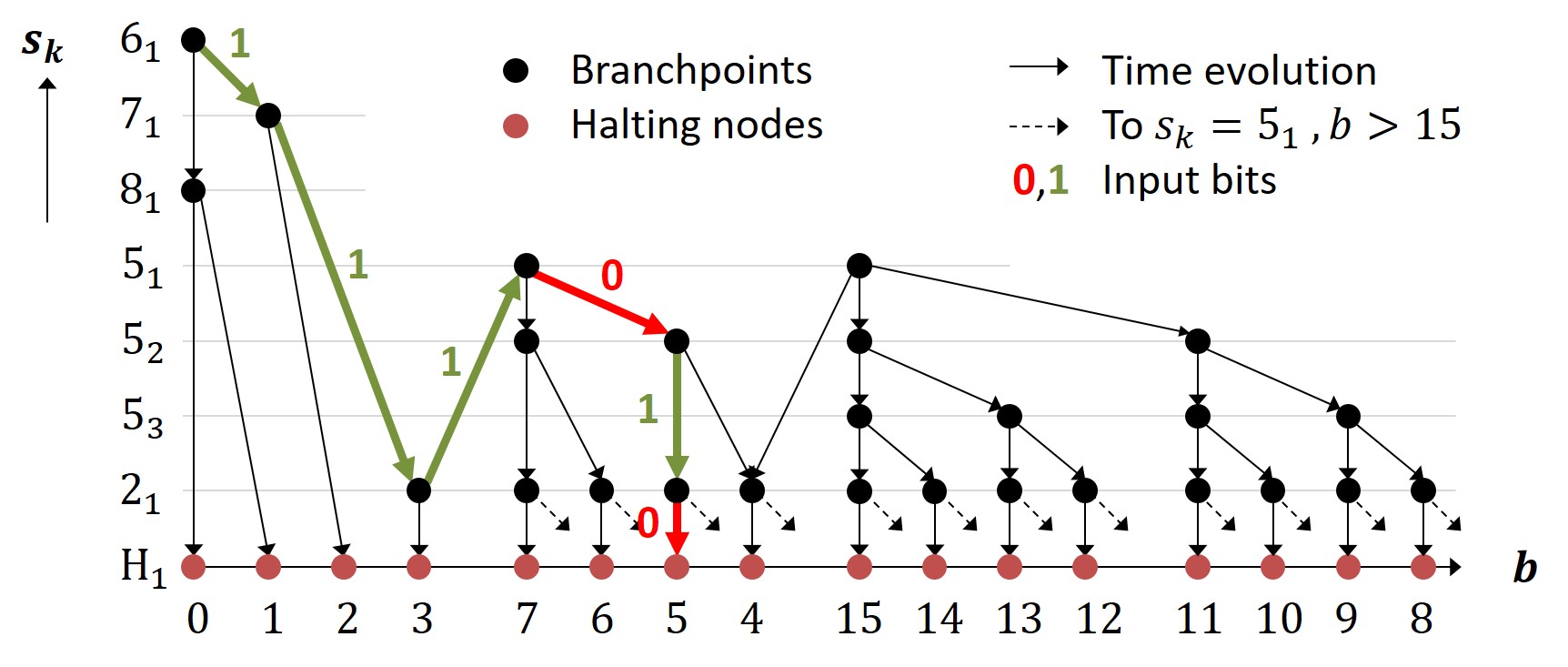} 
	\caption{Paths in configuration space for the counting machine}
	\label{figC}
\end{figure}

Fig. 3 shows the computation paths and their first 15 output bit strings at the example of the counting machine. Fig. 3 can be thought of as an embedding of the tree graph of fig. 1 in configuration space. The bit string $b(\tau)$ is plotted on the horizontal axis, written as a decimal number $N$, except for the halting state "H", in which case $N$ denotes the length of the output bit string $1^N$. The states $s(\tau)$ and the position $k(\tau)$ of the work head are denoted by $s_k(\tau)$ and are plotted on the vertical axis (only certain configurations appear). The arrows show the direction of the computation; dashed arrows lead to configurations with $N>15$ and $s_k=5_1$. Black dots represent configurations before the paths split ("branchpoints"). The green/red bits shown along the example of the computation path leading to $B=11111_2$ ($N=5$) form the corresponding input program "111010". \\

It is instructive to compare with a particle on a Euclidean $N$-dimensional lattice with coordinates $\vec x$. The amplitude for the particle to go from point $\vec a$ to $\vec b$ is (see, e.g., \cite{poly}):
\begin{eqnarray}
G(\vec a,\vec b)&=&\sum_{l=0}^\infty N_l(\vec a,\vec b)\cdot \exp\{-\beta\cdot l\},
\end{eqnarray}
where $N_l(\vec a,\vec b)$ is the number of paths $\vec x(\tau)$ (or "world-lines") of length $l$ on the lattice (measured in lattice spacings) that interpolate between $\vec x(0)=\vec a$ and $\vec x(l)=\vec b$. As $l\rightarrow\infty$, $N_l$ diverges as $\exp\{\beta_c\cdot l\}$, where the precise value of $\beta_c$ depends on the lattice. By setting
$$\beta =\beta_c+\epsilon m^2\ \ \text{as}\ \ \epsilon\rightarrow 0 ,$$
a continuum limit can be reached, where the model describes a lattice-independent relativistic particle of mass $m$ in $N$-dimensional Euclidean space with restored Poincar\'e symmetry. After a Wick rotation, this becomes a quantum mechanical particle in Minkowski space. \\

At least formally, the (high-dimensional) configuration space of the Turing machine is analogous to the lattice $\vec x$ of the particle, and the computation histories are analogous to the world lines $\vec x(\tau)$. The input bits on the program tape, which "live" on the paths in fig. 3, correspond to the random variables $\vec\eta(\tau+1)=\vec x(\tau+1)-\vec x(\tau)$, which "live" on the particle's world line and describe its incremental movement on the lattice. Of course, a difference is that the computation of the Turing machine is irreversible, following the arrows in fig. 3, while the motion of the particle is reversible and symmetric: $G(\vec a,\vec b)=G(\vec b,\vec a)$. \\

Both for the particle and for the computation, the average length of the world lines diverges for $\epsilon\rightarrow 0$. The key question is
what the computational analogue of the continuum limit of the relativistic quantum particle is. In trying to shed light on this, let us next discuss the corresponding Hilbert space in the case of the computation. 

Using the classification of \cite{fort}, let us begin by regarding the ensemble (\ref{B}) as a probabilistic Turing machine. It acts on a Hilbert space of states, which are superpositions of the individual configurations $S$, each of them occurring with probability $p_S$:
\begin{equation}
\vert\Psi_\tau\rangle=\sum_S\ \psi_S(\tau) \vert S\rangle\ \ \ \text{with}\ \ \ \sum_S p_S=\sum_S\vert \psi_S\vert^2\le1,\label{psi}
\end{equation}
 Given a pure state $\vert S\rangle$ at time $\tau$, the next state at time $(\tau+1)$ is a superposition of the two possible successor configurations, each of which occurs with probability $e^{-\beta_c}\sim1/2$. E.g., for the counting machine, if we define $l_N$ as in (\ref{J}) and $\psi_B({\tau+1}) = \psi_B(\tau)$ for $\tau\ge l$ (i.e., after the computation halts), then, in the limit $\tau\rightarrow\infty$, the output state  converges to
$$\vert \Psi_\infty\rangle=\sum_{N=0}^{\infty}\psi_N \vert N\rangle \ \ \text{with}\ \ \ \vert N\rangle\equiv\vert 1^N,\text{Halt},1\rangle, \ \ \psi_N=2^{-l_N/2}\ \Rightarrow\ \sum_N\vert\psi_N\vert^2 =1.$$
The evolution of $\vert \Psi_\tau\rangle$ from time $\tau$ to $\tau+1$ can be described by a transfer matrix $T$. This evolution is generally not reversible, and for a universal Turing machine not even probability is conserved due to non-halting programs (hence the $\le$ sign in (\ref{psi})). Thus, the transfer matrix $T$ is not unitary and the computation generates heat. However, $T$ can be extended such that the time evolution converges to a unitary one within the subspace $\hat S$ spanned by output states $\vert B\rangle$. E.g., an arbitrary phase $\delta_B$ can be added in the relation 
$$\psi_B({\tau+1})=e^{i\delta_B} \psi_B(\tau)\ \ \text{for}\ \ \tau\ge l.$$
An example with applications to error-correction codes is Manin's definition $\delta_B=K(B)$ \cite{Y2}. It will be interesting to explore other applications for more general Transfer matrices $T=\exp\{i\tau H\}$, with a self-adjoint Hamiltonian $H$ acting on the space $\hat S$ of halting states. This turns the probabilistic Turing machine into a quantum Turing machine acting on $\hat S$.\\ 

In appendix A6, we consider the ensemble (\ref{B21}) for output strings $B_L$ of large but fixed length $L$. Using the above formalism, we discuss the ground state of the system for a (non-universal) toy model of a Turing machine, the "kink machine". At low-temperature ($\epsilon=\beta-\beta_c\gg1/\lg L$), there is a degenerate, ordered ground state. As the temperature increases, kinks start to appear at $\epsilon\sim1/\lg L$. They become dense at the critical point, thereby destroying the long-range order. A similar condensation of defects in spin models typically restores spontaneously broken symmetries and accompanies second-order phase transitions, where the models are described by universal continuum field theories. We must leave it for future work to examine whether analogous phenomena occur for a UTM.

\section{Outlook} 

To briefly summarize our results, let us denote programs that produce a given output bit string $B_L$ of length $L$ by $p(B_L)$, and decompose our Gibbs ensemble (\ref{B}) as 
\begin{eqnarray}
Z(\beta)&=&\sum_L Z_L(\beta)\ \ \ \text{with} \label{out1} \\
Z_L(\beta)&=&\sum_{B_L}\ e^{-\beta\cdot K_\beta(B_L)}\ \ ,\ \  e^{-\beta\cdot K_\beta(B_L)}= \sum_{p(B_L)} e^{-\beta\cdot l(p(B_L))} .\label{out2} 
\end{eqnarray}\

We have shown that ensemble (\ref{out1}) has a first-order phase transition at the Chaitin point $\beta=\beta_c=\ln 2$. This phase transition is "almost" zeroth-order, in the sense that the free energy is continuous but almost discontinuous: it converges to its critical value, namely Chaitin's $\Omega$, more slowly than any computable function. However, for finite computation time and therefore limited $l$, this effective discontinuity is invisible, and one can smoothly interpolate between the classical limit $\beta\rightarrow\infty$ and the critical point $\beta=\beta_c$. 
\\

At the Chaitin point, we have shown that the latent heat (the average program size) and the average size of the output bit strings both diverge. This begs the question whether there is a continuum limit, in which long bit strings are described by continuum strings and the computation histories are represented by string world sheets. This must be decided in future work by adapting the tools of quantum mechanics and field theory to bit strings.\\

In section 9, we have taken a few preparatory steps in this direction. We have discussed analogies and differences with the continuum limit of a quantum particle on a lattice. We have also interpreted (\ref{out1}) as a probabilistic Turing machine and defined the analog of a Hilbert space of states. For the simplified toy model of appendix A6, we have examined the ground state of (\ref{out2}) for fixed length $L$. We have found a degenerate low-temperature ground state, whose long-range order is "washed out" near the critical temperature by kinks, similarly as in second-order phase transitions in spin models. \\

While the sum over $L$ in (\ref{out1}) yields a first-order transition for $Z(\beta)$, this raises the question whether a universal Turing machine has a second-order transition for $Z_L(\beta)$, when $L$ is held fixed. As precedents of statistical mechanical models of dynamic size $L$ that have (i) zeroth- or first-order phase transitions when one sums over $L$, and (ii) second-order transitions for fixed $L$, consider the Ising model or the sine-Gordon model (referred to as "matter") on a random lattice \cite{M,moore,I}. In essence, the partition function is of the form
$$Z(\beta)=\sum_L \exp\{-\beta L\} \cdot Z_\text{matter}(L)\ \  ,\ \ L=\text{number of lattice sites}.$$
These models have zeroth-order phase transitions in $\beta$. However, as $\beta\rightarrow\beta_c$, $\langle L\rangle$ diverges, and the models have continuum limits, where the "matter" undergoes second-order phase transitions on continuous random surfaces, whose area grows with $L$. These systems are described by two-dimensional field theories known as "noncritical string theories".\\

Why should we care about the order of the phase transition of the ensemble (\ref{B})? At second-order phase transitions, statistical mechanical models that are very complex on a microscopic scale often have simple, universal macroscopic descriptions in terms of continuum field theories. E.g., water and steam at a temperature of 374$^\text{o}$C and a pressure of 218 atm is described by a scalar $\phi^4$ theory, which can be used to compute its critical exponents. At this critical point, almost all microscopic details become irrelevant. E.g., exactly the same critical exponents describe the second-order phase transition of CO$_2$, of simple lattice gas models, and of any other system in the same universality class.\\

Ensemble (\ref{out1},\ref{out2}) certainly looks intractable on a microscopic scale: the action is higly nonlocal, and not even computable by any halting program. Our hope of better understanding Chaitin's Omega therefore rests on the hypothesis that the ensemble has a tractable continuum limit at the Chaitin point, in the sense that it can be modelled by what could be called a "logical quantum field theory" that is independent of the algorithmic details. We could then examine the ground state by approaching the critical temperature from below, as exemplified at the simple toy model of appendix A6. \\

Why do we want to better understand Chaitin's Omega in the first place? It provides a fascinating link between information theory and statistical mechanics, for which many potential applications can be foreseen. Apart from the envisioned simulation of strings and the examination of their ground states, those include the analysis of heat generation in quantum Turing machines, the simulation of the evolution of species in terms of a Turing machine acting on DNA strings, or the generalization of the unitary time evolution of \cite{Y2} to additional applications beyond error-correcting codes and Zipf's law. 

\section*{Acknowledgements} 

I would like to thank my brother Juergen Schmidhuber for arising my interest in information theory. The current work was inspired by his idea of the "Great Programmer" \cite{J}. This research is supported in part by grant no. CRSK-2 190659 from the Swiss National Science Foundation.\\

\newpage
\section*{Appendix A: Review of Turing Machines}

The appendix is organized as follows. A1 presents a Turing machine that contains only a work tape and no program tape. An example is given in A2. In A3, Chaitin's definition of a Turing machine is recalled. Using the example of A2 as a building block, we realize the counting machine of section 4 in A4. A5 contains a supplementary argument to section 5. A6 constructs a computable transfer matrix/Hamiltonian for our ensemble (\ref{B}).

\subsection*{A1. A simple Turing Machine}

Our first example of a Turing machine contains a "work tape" that extends infinitely in both directions. It consists of cells that are blank, except for a finite, contingent bit string of 0's and 1's (the "input string"). A blank cannot be written between 0's or 1's, so it is not equivalent to a third letter in addition to 0 and 1. Rather, blank areas mark the beginning and end of the string on the work tape. On the first cell of the input string sits a head, which can read, write, and move in both directions. The head can be in one of several states, labelled by 1, 2, 3, ... ,H. At each step, the machine operates as follows:

\begin{enumerate}\addtolength{\itemsep}{-5 pt} 
	\item it reads the bit on the work tape on which the head sits ($0$, $1$ or a blank)
	\item depending on that bit and on its internal state, it writes a $0$, $1$ or a blank in that cell on the work tape. It may only write a blank if the cell has a blank neighbour, to ensure that the binary string remains contingent
	\item it moves the head either one cell to the left or one cell to the right
	\item it may or may not change its internal state
	\item If and when it reaches the state "H", it halts
\end{enumerate}

\subsection*{A2. An example}

As an example, consider a Turing machine with 5 states $1,2,3,4,H$. The first six columns of table 1 define how this particular machine writes $0,1$ or $2$ ($2$ denoting a blank), then moves left ($-1$) or right ($+1$), and then switches to a new state, depending on the input bit it reads (left column) and the state it is in (top row). \\

\begin{figure}[h]\centering
	\includegraphics[height=5cm]{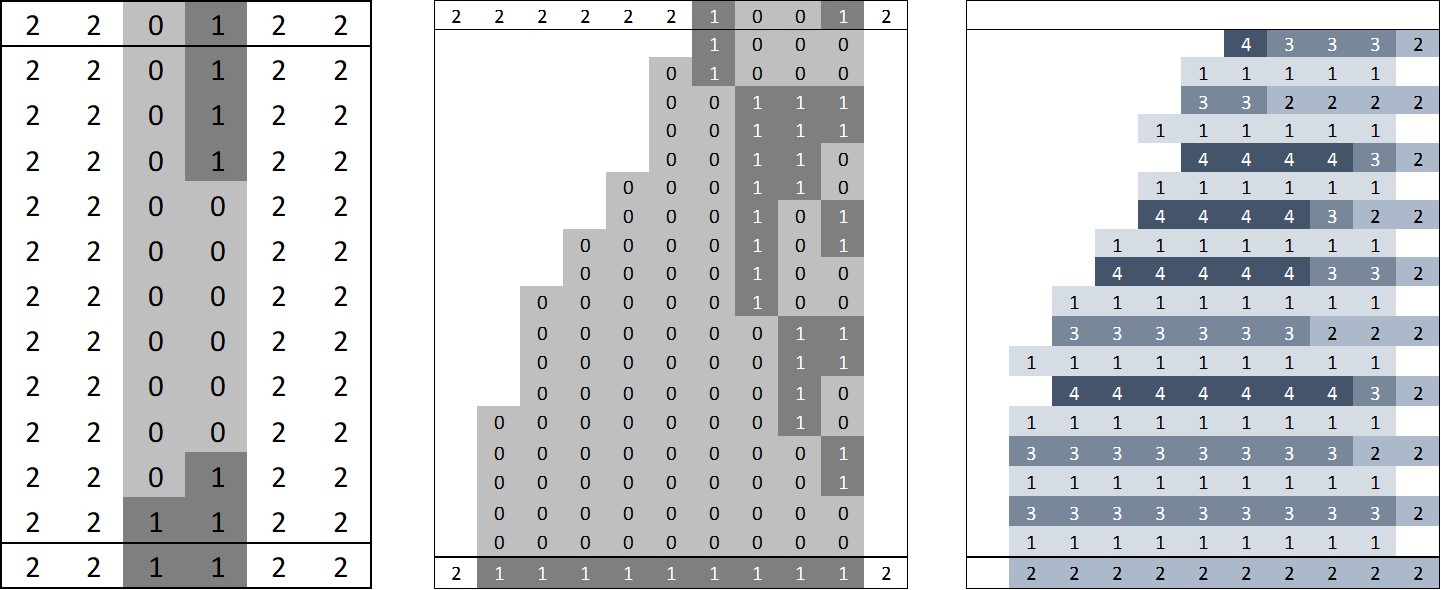}\ \ \ \ 
	\caption{A simple Turing machine}
	\label{figD}
\end{figure}

First, let the input string be "01". Fig \ref{figD} (left) shows a two-dimensional graph of the evolution of the bit string, with a new row appended for each time step. The machine
\begin{itemize}\addtolength{\itemsep}{-5 pt} 
	\item starts in state $1$ on the first bit of the work tape, which is $0$. 
	\item writes a $0$, remains in state $1$ and moves right to the next bit, whose value is $1$. 
	\item writes a $1$, remains in state $1$, and moves right to the next bit, which is blank. 
	\item switches to state $2$, and moves back left to the prevous bit, whose value is $1$. 
	\item overwrites it with $0$, switches to state $3$, and moves left, and so on. \\
\end{itemize}

\begin{tabular}{ |l|c||c|c|c|c||c c|c c|c c|c c c|}	\hline
	\multicolumn{2}{|l||}{{\it Table 1} } &\multicolumn{4}{|c||}{Current state} &\multicolumn{9}{|c|}{Current state \& program bit} \\ \hline
	\bf Operation &Work Bit& 1  & 2 & 3 & 4& $5_0$&\bf $5_1$&\bf $6_0$  &\bf $6_1$ &\bf $7_0$ &\bf $7_1$&\bf $8_0$ &\bf $8_1$ &\bf $8_2$\\	\hline\hline
	Write&0 	&0&1&0&0  	&-&-	&2&1&1&1&2&1&2\\  
	on work&1 	&1&0&1&1 	&0&1	&2&1&1&1&2&1&2\\  	
	tape&2 		&2&2&2&0	&2&2	&2&1&1&1&2&1&2\\  \hline	
	Set &0 		&1&2&3&4	&-&-	&8&7&H&1&H&H&6 \\	
	the new &1 	&1&3&4&4	&5&5	&8&7&H&1&H&H&6 \\	
	state &2 	&2&H* &1&1	&2&2	&8&7&H&1&H&H&6\\	\hline
	Move &0 	&1&-1&-1&-1	&-&-	&1&1&-1&-1&-1&-1&1\\
	on work&1  	&1&-1&-1&-1	&1&1	&1&1&-1&-1&-1&-1&1\\
	tape&2 		&-1&1&1&1	&-1&-1	&1&1&-1&-1&-1&-1&1\\ \hline \hline
	Move on&0 	&-&-&-&-	&-&-	&1&1&0&1&0&0&1\\ 
	program&1 	&-&-&-&-	&1&1	&1&1&0&1&0&0&1\\ 
	tape&2 		&-&-&-&-	&0&0	&1&1&0&1&0&0&1\\ \hline
\end{tabular}\\  \\

At some point, the machine lands on a blank bit in state $2$, moves one bit to the right and halts. The output string are two $1$'s in a row. By modifying the input string, other output strings can be produced. The reader may verify that if the input string is the number $b$ in binary code, then the output string of this particular Turing machine always consists of $b$ $1$'s in a row, with the head halting on the first cell with value "1".\\

Fig. \ref{figD} (center) shows this computation in condensed form for $b=1001_2=9_{10}$. By "condensed", we mean that each time step now corresponds to a new square, rather than a new row, such that the computation time is the area of the graph. The head of the machine moves along the rows of the graph, and each time it changes direction, a new row is appended. For completeness, fig. \ref{figD} (right) also shows the state of the machine at each point in the computation. The machine moves right along the light grey rows (state 1) and left along the other (blue) rows (states 2,3,4).   We call these graphs "bit string world sheets".
\\

As an example of a universal Turing Machine (UTM) that can simulate all other Turing machines, consider our brain: given the above table for any Turing machine, we can read it and use it to simulate the machine as you have just done if you have followed the exercise. Essentially, the table becomes part of the input, rather than being hard-coded into the Turing machine. For a more specific example of a universal Turing machine, see, e.g., \cite{minsky}. A UTM is arbitrarily flexible and can quickly compute strings with one Turing machine that take a long time or are impossible to compute with another machine.\\

There are many alternative, but equivalent definitions of Turing machines. E.g., one can introduce other symbols in addition to $0$ and $1$, or more states, or one can work with several parallel work tapes instead of just one. 

\subsection*{A3. Chaitin's Machine}

In Chaitin's definition, there is a read-only "program tape" of finite length, in addition to the work tape. The program tape begins with a blank cell followed by a finite bit string of $0$'s and $1$'s, the "program". On the program tape sits another head, the "program head". Initially, it sits on the blank cell. At each step, the machine performs the following operations in addition to steps 1-5 of subsection A1:

\begin{itemize}\addtolength{\itemsep}{-5 pt} 
	\item initial step: it reads the bit on the program tape on which the head sits
	\item last step: it moves the program head either one cell to the left or leaves it where it is
\end{itemize}

The machine either halts or runs forever without reading any more program bits. As a result, the set of input programs, from the first to the last bit that has been read by the machine, is prefix-free.

\subsection*{A4. The Counting Machine}

As an example within Chaitin's framework, we present an implementation of the counting machine of section 4. We begin with the Turing machine of appendix (A1), and add a finite read-only program tape, on which the programs of section 4 are written. We start with a work tape that is initially blank.  \\

We first add three additional states $6$, $7$, $8$, whose role is to read the first two bits on the program tape and get the machine started (steps 1 and 2 of section 4). The operations in states $6$, $7$, $8$ depend only on the program bit on which the program head sits, and not on the work bit on which the work head sits. They are defined in table 1. The machine is initially in state 8 (in states 6 and 7, the program bit is then never 2).  \\

Next, we slightly modify state $2$ in table 1 as follows: if the machine is in state $2$, and the head on the work tape sits on a blank, then it switches to state $H$ {\it only} if the head of the program tape sits on a $0$. Otherwise, it moves to a new state 5 (the asterix in H* indicates that $H$ is replaced by 5, if the program bit is 1). The operations of the new state $5$ are also defined in table 1. Its role is to write a new portion from the program tape onto the work tape, thereby over-writing the contingent sequence of 1's. Its operations depend both on the current work bit and on the current program bit (the machine is never in state 5 when the program head is on a blank or when the work head is on a 0). It is straightforward to verify that this machine indeed represents the counting machine of section 4.

\subsection*{A5. A Supplementary Argument}

\begin{figure}[h]\centering
	\includegraphics[height=4.2cm]{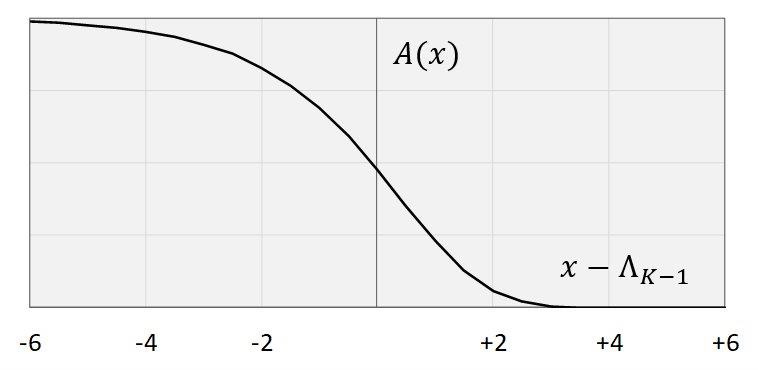}\  
	\caption{The function $A(x)$}
	\label{figE}
\end{figure}

In section 5, we want to evaluate the $K$-th part of the partition function
\begin{eqnarray}
\hat Z_K(\epsilon)= \sum_{n_2,...,n_{K}} 2^{-(6+n_2+...+n_{K-1})}\cdot {1\over2}\ e^{-\epsilon n_K}\ \ \ \text{in the case}\ \ \ {1\over\epsilon}=\Lambda_K
\end{eqnarray}
where $K\ge4$, $n_2$ runs from 4 to 7, $n_{i+1}$ runs from $2^{n_i-1}$ to $2^{n_i}-1$, and $\Lambda_K$ is the largest possible value of $n_K$. Specifically, $\Lambda_3=127, \Lambda_4=2^{127}-1$, and therefore $\Lambda_{K-1}\sim\lg(\Lambda_K)=\lg(1/\epsilon)$ to high accuracy for $K\ge4$. Defining $M=2^{n_{K-1}}$, the sum over $n_K$ yields
\begin{eqnarray}
\hat Z_K(\epsilon)&=&\sum_{n_1,n_2,...,n_{K-1}} 2^{-(6+n_2+...+n_{K-2})}\cdot A(n_{K-1},\epsilon)\\
A(n_{K-1},\epsilon) &=&{1\over 2M}\sum_{n_K=M/2}^{M-1}   e^{-\epsilon\cdot n_K}
={1\over{2 M\epsilon}}({e^{-M\epsilon/2}-e^{-M\epsilon}})  \label{E}
\end{eqnarray}
$A(x,\epsilon)$ is plotted in fig. \ref{figE}. It is a monotonously decaying function with
\begin{equation}
A(x,\epsilon)\rightarrow \left\{ \begin{array}{l}
{1\over 4}\ \ \ \text{for}\ \ \  x \ll \lg {1\over\epsilon}\sim\Lambda_{K-1}\\
0\ \ \ \text{for}\ \ \ x\gg \lg {1\over\epsilon}\sim\Lambda_{K-1} \end{array}\right. 
\end{equation}

$n_{K-1}$ runs from $2^{n_{K-2}-1}$ to $2^{n_{K-2}}-1$. Only for the maximal value of $n_{K-2}$ are there a few values of $n_{K-1}$ near $\Lambda_{K-1}$, for which $A$ differs significantly from $1/4$. Even in this case, the contribution of these differences is 

\begin{itemize}\addtolength{\itemsep}{-5 pt} 
\item 
small (of order 1\%) for $K=4$: for the highest value $n_2=7$, $n_3$ runs from 64 to 127. Only the last few of these $n_3$ contribute significantly to the difference
\item 
practically zero for $K\ge 5$: e.g., for $K=5$ and the highest value $n_3=127$, $n_4$ runs from $2^{126}$ to $2^{127}-1$. Only a tiny portion of these $n_4$ contribute to the difference
\end{itemize}

As long as $K\ge4$, we can thus approximate $A$ by ${1/4}$ for $x \le \Lambda_{K-1}$ to obtain $\hat Z_K=2^{-K-3}$ as claimed in section 5. An analogous argument, not repeated here, shows that we can approximate $A$ by $0$ for $x > \Lambda_{K-1}$ to obtain $\hat Z_{K+1}=0$, as long as $K\ge4$. 

\subsection*{A6. The Kink Machine}

The counting machine of section 5 produces only trivial output strings that consist of $L$ 1's in a row. In this appendix, we define another non-universal Turing machine, which we call the "kink machine". It acts on these output strings $1^L$ as follows:

\begin{enumerate}
\item The kink machine first simulates the counting machine and reads bits from the program tape to produce another integer $M\le L$. It then reads the next $M$ bits from the input tape, which form a bit string $m$ of length $\vert m\vert =M$. The machine then overwrites the output $1^L$ of the original counting machine by repeated copies of $m$. E.g., if $L=20$, $M=3$, and $m$="101', the output is (using (\ref{J})):
$$B=10110110110110110110\ \ \ \text{with weight}\ \ \ 2^{-l_M-M}={1\over32}.$$ 
\item The machine then simulates the counting machine again to produce another number $K\le L$ (where $K$ may be zero). It then creates $K$ "kinks" at positions $N_1, ..., N_K$. A kink at $N_1$ means that $N_1$-th bit and all subsequent bits of $B$ are flipped from 0 to 1, resp. from 1 to 0. E.g., a single kink at position $N_1=8$ acting on $B$ produces
$$B'=1011011{\color{darkred}1001001001001}\ \ \ \text{with weight}\ \ \ {1\over32}\cdot 2^{-2-\lg L}={1\over32}\cdot {1\over4 L},$$ 
because it takes ($\lg L$) bits to specify the position $N_1\in\{1, ..., L\}$, and we have used $l_K=2$ for $K=1$ from (\ref{J}). Then the kink machine halts. 
\end{enumerate}
We now let the kink machine act on "$1^L$" to produce strings of fixed large length $L$. To be specific, $L\sim2^{32}$ may be the size of the human genome in bits. The final output state is
$$\vert \Psi_\infty\rangle=\sum_{B}\psi_{B} \vert B\rangle\ \ \ \text{with}\ \ \ \psi_{B}=2^{-l_B/2}\cdot e^{-\epsilon\cdot l_B/2} \ ,  \ l_B=l_M+M+l_K+K\cdot\lg L,$$
as long as $K\ll L$ (when $K$ is of order $L$, the complexity is lower. E.g., 2 kinks at the same position are equivalent to no kink).\\ 

Let us gradually take $\beta=\beta_c+\epsilon\rightarrow\beta_c$, starting with very low-temperature ($\beta\rightarrow\infty$).  From (\ref{J}),  there we have a degenerate ground state (with $m=1,0,10,01$ with $l_1=l_2=2$):
$$\vert \Psi_\infty\rangle\sim{1\over2}\Big(\vert 1111...11 \rangle+\vert 0000...00\rangle+\vert 1010...10\rangle+\vert 0101...01\rangle \Big).$$

It is somewhat analogous to the low temperature ground state of the Ising model, which consists of spins that either all point up ($\uparrow...\uparrow$) or all point down ($\downarrow...\downarrow$). In our case there is no spontaneous symmetry breaking, as it takes only a short program (i.e., finite energy) to convert the 4 components of the ground state into each other. Still we can interpret the 2-tuple $(M,m)$ as an order parameter, analogous to the magnetization in spin models.  \\

We now increase the temperature such that $1/\lg L \ll \epsilon\ll1$. E.g., for $L\sim 2^{32}$ we may choose $\epsilon=1/10$. Then states with kinks (i.e., $K>0$) are still suppressed, and only $B_m$ with short $m$ contribute significantly to the ground state 
$$\vert \Psi_1\rangle=\sum_m\psi_m  \vert B_m\rangle\ \text{with}\ \ \psi_m\sim e^{-(l_M+M)/2}\ \ \text{where } M=\vert m\vert\ll\lg L.$$
Although the simple correlation between spins at different positions on the bit string is generally zero, this is still a low-complexity state with long-range order (a variant of the correlation that measures this long-range order can be defined, but we omit this here). \\

There are $L$ possible values of the position $N_K$ of a kink, and it takes a program of length $\lg L$ to specify $N_K$. Thus, for $K\ll L$, the probability of $K$ kinks is proportional to $2^{-l_K}\exp\{-K\epsilon\lg L\}$. As we decrease $\epsilon$ below $(1/\lg L)$ (say, $\epsilon=0.01$ for $L\sim 2^{32}$), kinks start to appear in the ground state. As we approach the critical point ($\epsilon\rightarrow0$), these kinks "condense" in the sense that the average distance between kinks (i.e., between the values of $N_K$) becomes of order 1. This destroys the long-range order of the ground state. \\

Note that there is a significant difference to the one-dimensional Ising model, for which the energy of a kink is independent of $L$, and thus kinks condense at any temperature. For the kink machine, they only condense at the critical temperature. In this respect, our model is more similar to the two-dimensional Ising model, where defects (Peierls droplets) condense near the critical temperature. In the latter case, there is a second-order phase transtion, where the order parameter (the magnetization) goes to zero, the correlation length diverges, and the system shows a characteristic scaling behavior. \\

Of course, the kink machine is much more primitive than a universal Turing machine. We must leave it for future work to examine whether a similar second-order phase transition as in spin models occurs for a universal Turing machine.

\end{document}